\documentclass[fleqn,usenatbib]{mnras}

\usepackage[T1]{fontenc}

\DeclareRobustCommand{\VAN}[3]{#2}
\let\VANthebibliography\thebibliography
\def\thebibliography{\DeclareRobustCommand{\VAN}[3]{##3}\VANthebibliography}

\usepackage{graphicx}	
\usepackage{amsmath}	
\usepackage{amssymb}	

\usepackage[english]{babel}

\newcommand{\teff}{\ensuremath{T_\mathrm{eff}}}

\newcommand{\logg}{\ensuremath{\log g}}
\newcommand{\kms}{$\rm km\,s ^{-1}$}

\title[]{Chemical Composition of a Palomar 12 Blue Straggler}

\author[L. Pasquini et al.]{
L. Pasquini,$^{1}$\thanks{E-mail: lpasquin@eso.org, Observations based on ESO observing program 60.A-9505(A)}
P.Bonifacio,$^{2}$
L. Pulone,$^{3}$
A. Modigliani,$^{1}$
E. Brocato,$^{3,4}$
L. Sbordone,$^{5}$
S. Randich,$^{6}$
and G. Cupani $^{7}$
\\
$^{1}$  ESO - European Southern Observatory, Karl-Schwarzchild-Strasse 2, 85748 Garching bei M\"{u}nchen, Germany \\
$^{2}$ GEPI, Observatoire de Paris, Universit\'e PSL, CNRS, Place Jules Janssen, F-92195 Meudon, France \\
$^{3}$ INAF - Osservatorio Astronomico di Roma, Via di Frascati 33, I-00078, Monte Porzio Catone (RM), Italy \\
$^{4}$ INAF - Osservatorio Astronomico d’Abruzzo, Via M. Maggini snc, I-64100 Teramo, Italy \\
$^{5}$ ESO-European Southern Observatory, Alonso de Cordova 3107, Vitacura, Santiago, Chile \\
$^{6}$  INAF-Osservatorio di Arcetri, Largo E. Fermi 5, Firenze, Italy \\
$^{7}$ INAF - Osservatorio Astronomico di Trieste, Via Giambattista Tiepolo 11, 34131 Trieste 
}

\date{Accepted XXX. Received YYY; in original form ZZZ}

\pubyear{2022}

\begin{document}
\label{firstpage}
\pagerange{\pageref{firstpage}--\pageref{lastpage}}
\maketitle

\begin{abstract}
With the equivalent area of a 16m telescope, ESPRESSO in 4UT mode allows to inaugurate  high resolution spectroscopy for solar-type stars belonging to  extragalactic globular clusters.
 We determine the chemical composition of an extragalactic blue straggler. The star has a G magnitude of 19.01 and belongs  to the globular cluster Pal12, that is associated  to the Sagittarius dwarf galaxy. 
Abundances are computed by using high resolution spectroscopy and LTE analysis.
Two  50 minutes ESPRESSO spectra, co-added, provide a Signal to Noise Ratio of $\sim$25 with a resolving power R$\sim$70000. This allows us to measure with good precision abundance of several (13) elements. Li could help to distinguish between formation models of  Blue Stragglers;  we are able to set a 3$\sigma$ upper limit of Li=3.1, which is still too high to discriminate between competing models. The abundances we retrieve for the BS are compatible with those of giant stars of Pal 12 published in literature, re-analyzed by us using the same procedure and line list. Small differences are present, that can be ascribed to NLTE effects, but for Mg the BS shows a large under-abundance. The most likely explanation is that the BS atmosphere is dominated by gas processed through the Mg-Al cycle, but we have no suitable Al or Na lines to confirm this hypothesis. 
We show that ESPRESSO with 4UT can be used to derive precise abundances for solar-type stars fainter than magnitude 19. At these magnitudes a proper sky subtraction is needed and in crowded field the targets must be chosen with outmost care, to avoid contamination of the sky fibre from nearby stars.
\end{abstract}




\section{Introduction}

The availability of the mode ``4 UT'' for the ESO ESPRESSO
high resolution spectrograph \citep{espresso}, by which the 
light of four  8m telescopes is fed to the spectrograph, opens the opportunity of high resolution spectroscopy of solar-type stars fainter than  magnitude 18, in a realistic short amount of observing time. 
In the context of the ``science verification'' observations of this mode, 
in which telescope time is awarded to several projects for a few hours,
we observed a Blue Straggler star in the Globular Cluster
Palomar 12 (hereafter Pal 12).  This cluster orbits
the Sgr dwarf spheroidal \citep{ibata}, is metal poor  ([Fe/H]$\sim$-1), and  about 30$\%$
younger than the classical GCs  \citep{pancino}. 

Just above  the Turnoff (TO) of globular cluster, as a blue extension, are located the Blue Stragglers. 
Blue Stragglers (BS) have been for years enigmatic objects, and still now 
their formation mechanism is not fully 
understood. While, for instance, it is clear that they are the product of a merging,  
their general abundance pattern is the same as main sequence stars of the  cluster they belong to
\citep{SS2000}
and it is debated whether they form as result  of collisions between stars in clusters or as a result of mass transfer between, or merger of, the components of  short-period binaries.  
 Depending on the formation mechanism, for instance their Li abundance is expected to be around A(Li)=1.0 
\citep{Glebbeek}.
Beside the scientific interest of determining the BS abundances, this work is relevant as a 
 pilot study, to test to which extent it is  feasible to derive reliable abundances at these 
 faint magnitudes. 


\section{Data}
\label{data}

The star has been observed with the ESPRESSO spectrograph at the VLT telescope \citep{espresso} 
 in  4UT-mode. 
 In this mode, all the four 8 meter VLT Unit Telescopes observe the same object and convey the light to the 
 ESPRESSO spectrograph,  that is fed by a 16m equivalent aperture telescope.  ESPRESSO in this mode provides a resolving  power of ~70000  and a rather heavy detector binning is needed to reduce the detector noise.  The simultaneous spectral coverage is between 380 and 780 nm  \citep{espresso}.
 
\begin{figure}
   \centering
   {\includegraphics[width=.4\textwidth]{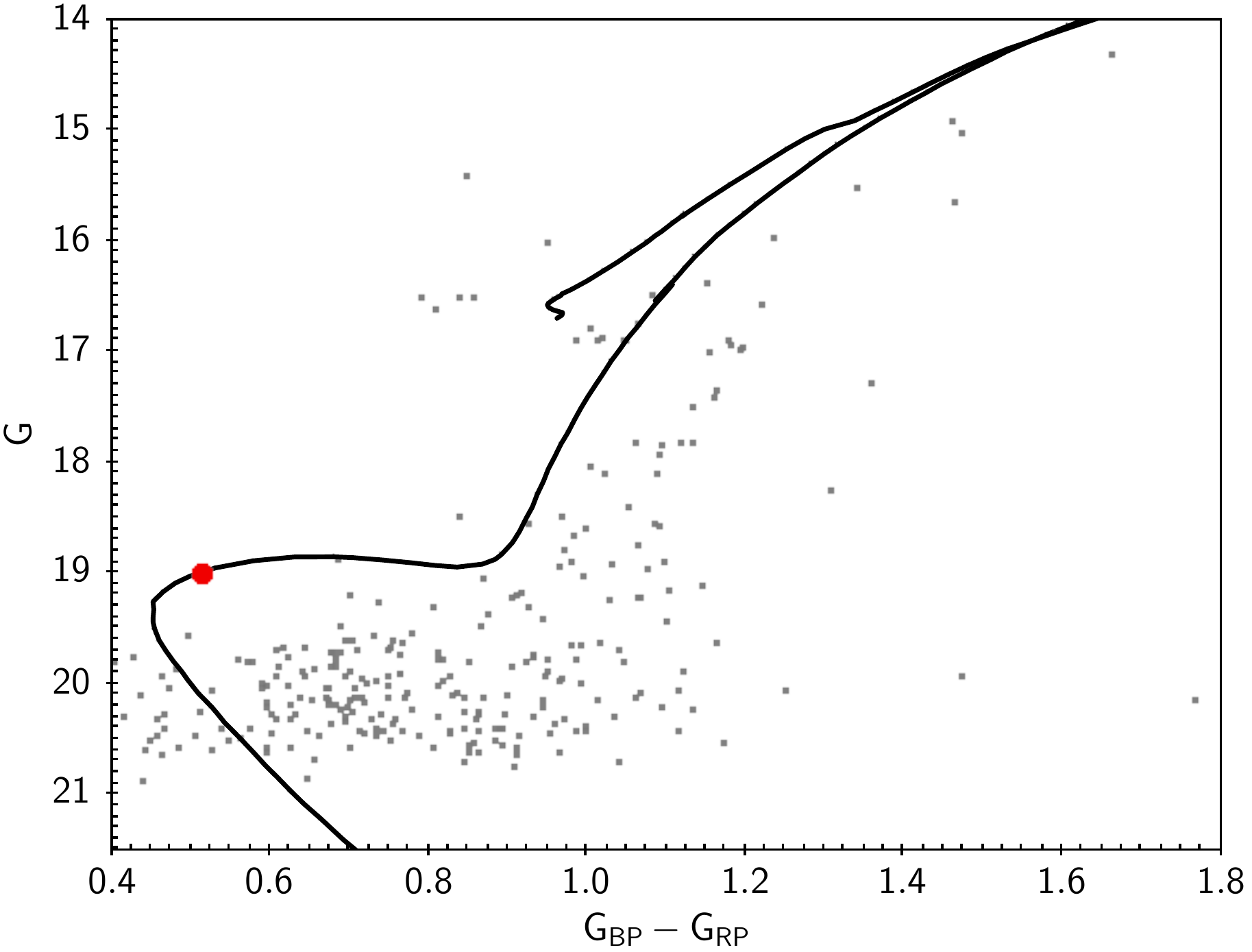}}
   \caption{Gaia colour-magnitude diagram for the stars within 5 arcminutes radius around Pal 12 center. Only stars with proper motions $<$5.5 milliarcesecond/yr have been retained.
The black line is a Parsec isochrone \citep{parsec} of age four Gyr, $Z/Z_\odot= 0.141$ and reddened for E(B-V)=0.02 mag. 
The absolute magnitudes of the ischrone have been shifted for a distance modulus of 16.47 \citep{pancino}.}
   \label{cmd}
   \end{figure}

The observed star is S1236 from the \citep{stetson89} list; it is a blue straggler, as it clearly appears from Figure \ref{cmd} showing the GAIA Colour Magnitude diagram containing all the stars within 5 arcminutes from the cluster center and with proper motions $<$ 5.5 milliarcsec/yr. The selected star is enclosed in the red square (GAIA EDR3:  source ID=6817951001157440512 , ra=326.6583556251155 , dec=21.259490392923386 G$_{mag}$=19.003, Bp$_{mag}$-
 Rp$_{mag}$=0.477, Bp$_{mag}$-G$_{mag}$=0.117.).
 
\begin{figure}
  \centering
  {\includegraphics[width=.4\textwidth]{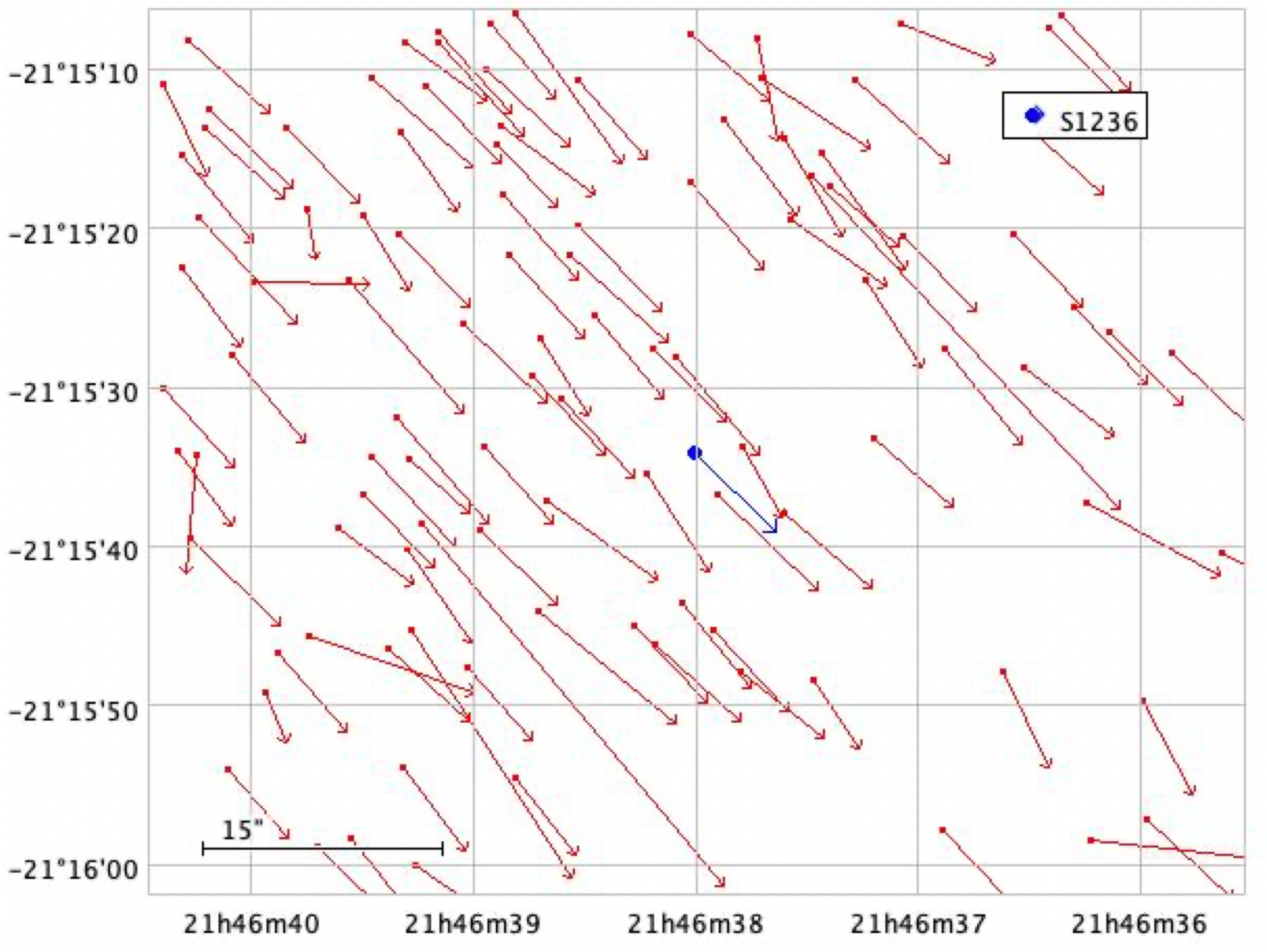}}
  \caption{Proper motions from the GAIA DR3 release for 532 stars in a radius of five arcmin  around the Pal 12 center. The proper motions of the observed blue straggler is marked  and is very close to most other stars, belonging to the cluster. }
   \label{inne_pm}
   \end{figure}

 The star has a proper motion of -3.54 and -3.50 milliarcseconds/yr (in $\alpha$  and $\delta$ respectively),  
 which is rather well coincident with the mean cluster  PM locus, as it appears from the analysis of the GAIA EDR3  $532$ stars located within a  $5\arcmin$ from the cluster center 
 The overall cluster stars proper motion, calculated from the 532 stars for which GAIA has a measure, is of 4.76 milliarcseconds/yr, while the star has 4.98, well within the 0.9 milliarcseconds/yr dispersion of the  cluster' stars. The Figure \ref{inne_pm} in 
 which the proper motions and positions of the stars at the very center of Pal 12 are represented, shows how our target shares the same velocity flux in module and direction as most of the other stars.
   
Three $\sim$55 minutes ESPRESSO spectra were acquired between September 2 and September 4 2019. Of  the two ESPRESSO fibres, one was devoted to the object, the second dedicated to record the sky.
The data have been reduced using the ESPRESSO pipeline.
The spectra we acquired with high detector binning (8x4 pixels), in order to minimize the impact of the detector Read Out Noise, and 
were reduced using the public version 1.3.2 of the ESPRESSO
data reduction pipeline and the ESOReflex workflow, with default parameters, optimised for this instrument setting (www.eso.org/pipelines). 

The ESPRESSO data reduction involves the following stages, executed as a chain:

-bias frames are processed to create a master frame and a determine the residuals, that is used in the recipes that process data with low signal to remove any residual structure present in the bias; similarly, master dark frames to obtain a map of the hot pixels;

-led flat frames (obtained illuminating the fibre with a led) are processed to map  non-linear pixels;
order tracing flats are used to trace the fibres location;

-flat fibre flats are used to obtain the fibre order profile (used to perform the optimal extraction), the order by order spectrum of the extracted flat (A and B) fibers, and  the blaze functions;

-contamination frames obtained with only one fibre illuminated by a calibration, to determine the  contamination between the two  fibres

-twilight spectra are used to compute the relative efficiency between the two fibres.

A three steps wavelength calibration (using Fabry-Perot etalon and Th-A lamp) follows as well as flux calibrations. 
Separate object and sky spectra are generated for each exposure.

 The sky spectra reveal that the third exposure was rather heavily contaminated by another stellar source, and at the end we have opted not to use this observation, because it was not possible to determine confidently the level of sky contamination. It has to be noted that for the other two exposures the sky counts are about 20 $\%$ of the object flux, a not negligible  quantity, and a wrong subtraction would  produce substantial errors in the measured  abundances, because it would deepen (or shallow) artificially the stellar absorption lines. 
In crowded fields the presence of nearby objects should be carefully checked  when observing with ESPRESSO, because the distance between the object and the sky fibres is fixed to seven arcseconds, and the sky orientation cannot be chosen. In addition, the  geometry of the projection on the sky is different for the four UT telescopes. When summing all of these effects the strong recommendation is to avoid targets having nearby stars around the 7 arcseconds distance, and  to this purpose a circle indicating the sky fibre radius has been now added to the observing preparation tool for ESPRESSO. 

\begin{figure}
   \centering
   {\includegraphics[width=.4\textwidth]{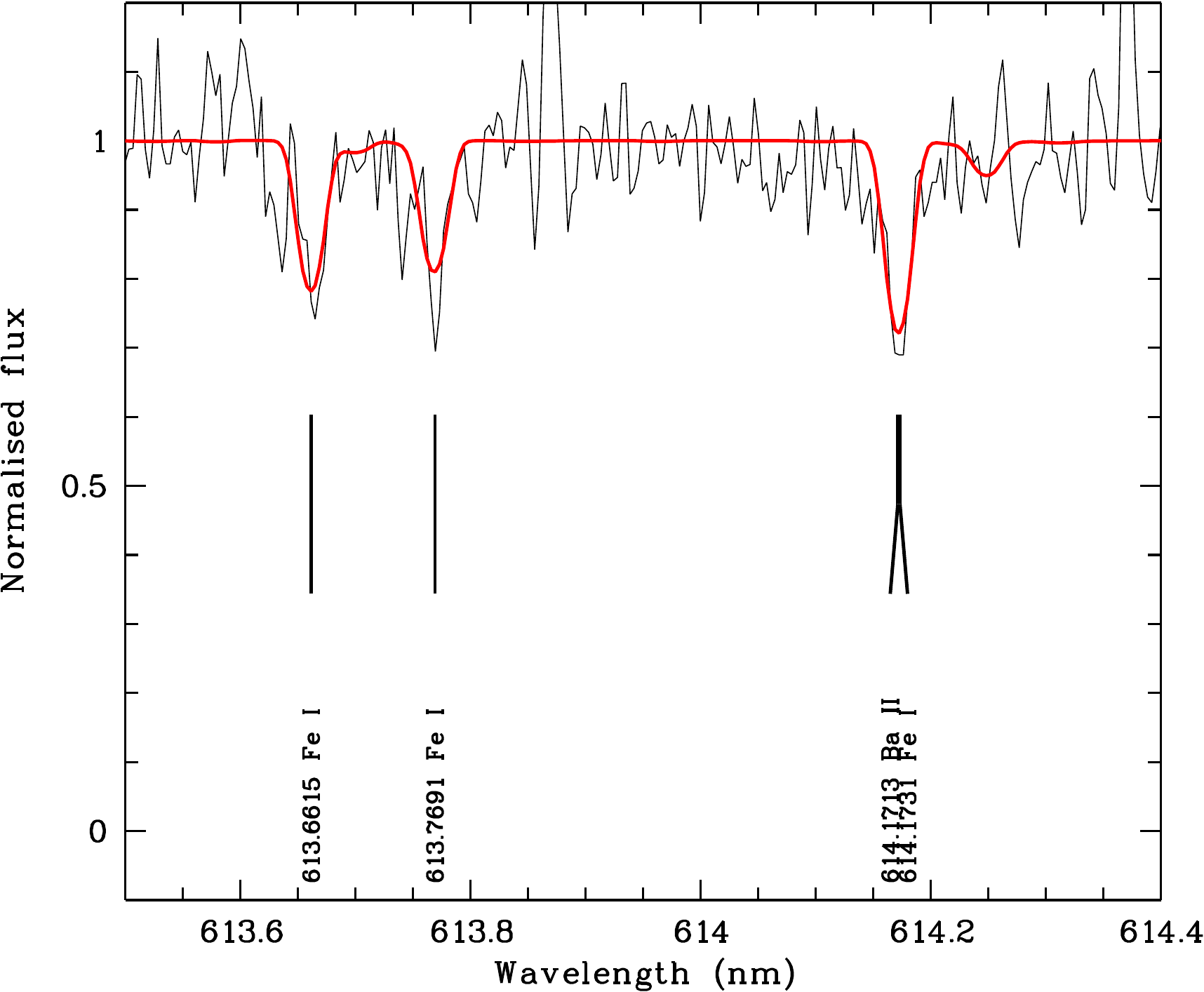}}
   \caption{Small portion of the ESPRESSO spectrum, with the line fit (in red) from MyGIsFOS \citep{SCBD14}. }
   \label{spectrum}
   \end{figure}

The typical signal level of the extracted spectra is of about 600 photoelectrons; the sky has been subtracted after passing an heavy smoothing, to avoid to decrease the S/N ratio of the stellar spectrum. The two 55 minutes spectra have been finally co-added, and the S/N ratio ranges  between 20 and 25 in the continuum in the visible and red portion of the spectrum.  
A portion of the final co-added spectrum is shown in Figure \ref{spectrum}, together with our fit. 

The stellar radial velocity (RV), as measured separately in our two exposures by fitting the photospheric lines, is of 27.98 $\pm$0.3 Km/s, in very good  agreement with the Pal 12 RV 
 of 28.5  km/sec, as measured by \citet{C04} on giant stars, confirming that the star belongs to the  cluster. 

\section{Atmospheric parameters, stellar rotation and chemical abundances} 

To determine the atmospheric parameters we used the Gaia EDR3 colour $G_{BP}-G_{BP}=0.477$  and a Parsec isochrone (Bressan et al. 2012) of age 3.66 Gyr and metallicity $Z/Z_\odot= 0.141$.

The isochrone was reddened for E(B-V) = 0.02, as appropriate for Pal 12 \citet{stetson89}.
By interpolating in the isochrone, for the observed $G_{BP}-G_{BP}$, 
we derived \teff = 7042\,K and log g = 3.95 (c.g.s. units).  Taking into account the
photometric uncertainty 0f 0.028 mag on the $G_{BP}-G_{BP}$ colour the error on \teff is 125\,K
and 0.07 dex on log g. 
Given the low S/N ratio we decided to keep fixed \teff and log g and to determine the
abundances from the analysis of the spectrum, see also \citep{MB2020} for
a discussion of the use of photometric versus spectroscopic atmospheric parameters.
We assumed a microturbulent velocity $v_t=1.5$\, \kms, as done by \citet{lovisi10} in their
study of Blue Stragglers in M\,4, as several of their stars
have parameters close to those of our target.  The S/N of the spectra
does not allow us to determine microturbulence in a  robust way,  especially because we can
measure too few lines on the linear part of the curve-of-growth.

The lines appear to be clearly too broad for the resolving
power delivered by the spectrograph. 
We attribute this to a 
non-negligible 
stellar rotation, which is observed in some blue stragglers \citep[see e.g.][and references therein]{lovisi10}.
In order to estimate this broadening we measured the Full Width Half Maximum (FWHM)
for a set of 73 \ion{Fe}{i} lines, selected to be not blendend
nor heavily saturated. The same set of lines was measured on synthetic
spectra that were rotationally broadened by, 6,8,10,12,14 \kms ~ and
then broadened with a gaussian of 4.28 \kms\ FWHM, to take into account
the instrumental profile. 
For each line we interpolated in the set of 5 values of FWHM measured
on the synthetic spectra to obtain one estimated value of rotational velocity. 
Before averaging the derived rotational velocities we removed all those
that would be extrapolating outside the interval 6-14 \kms. 
The mean rotational velocity so obtained is of 8.44 \kms $\pm 1.87$ \kms.

The spectral analysis has been carried on using the MyGIsFOS program \citep{SCBD14}
and a grid of synthetic spectra computed with SYNTHE an a grid of model atmospheres computed with ATLAS 9 \citep{K05}.
We used the atomic line list compiled by \citet{heiter},  molecular
lines from Bob Kurucz's site \footnote{\url{http://kurucz.harvard.edu/molecules/old/}}.
except for CH lines that were taken from \citet{masseron14}.
The grid spectra were individually rotationally broadened by 8.44 \kms
and the whole grid was broadened with a gaussian profile of 4.28 FWHM.  
Table 1 reports the abundance values retrieved. The line-to-line scatter for most elements having more than one measured line is compatible with the  S/N ratio of the spectrum. 
For Li we could only derive a three $\sigma$ upper limit of 23.7 m\AA, that corresponds to an upper limit of log (Li/H) +12 $<$ 3.1.
 
\begin{table}
\caption{Atmospheric abundances for the Star 1336\relax in Pal12 as obtained from our ESPRESSO spectra. 
Column 1: element, Column 2, number of lines used, column 3: abundance, Column 4: Uncertainty, Column 5: abundance scaled to solar }             
\label{abun}      
\centering  
\small
\begin{tabular}{l r r r r r }        
\hline\hline                 
Element & A(X)$_\odot$ &N(lines) & A(X) &  [X/H] & $\sigma$  \\    
\hline
\ion{Mg}{I}  &  7.54 & 3  &  6.20  &$-1.34$  &  0.13 \\
\ion{Si}{I}  &  7.52 & 1  &  6.85  &$-0.70$  & $-$  \\      
\ion{Ca}{I}  &  6.33 & 19 &  5.44  &$-0.89$  & 0.16 \\
\ion{Sc}{II} &  3.10 & 3  &  2.09  &$-1.01$  & 0.36  \\
\ion{Ti}{I}  &  4.90 & 5  &  4.26  &$-0.64$  & 0.23  \\  
\ion{Ti}{II} &  4.90 & 27 &  4.15  &$-0.75$  & 0.25 \\
\ion{Cr}{I}  &  5.64 & 9  &  4.84  &$-0.80$  & 0.31  \\
\ion{Cr}{II} &  5.64 & 5  &  4.57  &$-1.07$  &  0.28  \\
\ion{Mn}{I}  &  5.37 & 3  &  4.49  &$-0.88$  & 0.02 \\
\ion{Fe}{I}  &  7.52 & 136&  6.52  &$-1.00$  &  0.28\\
\ion{Fe}{II} &  7.52 & 21 &  6.35  &$-1.17$  &  0.23\\
\ion{Ni}{I}  &  6.23 &  6 &  5.24  &$-0.99$  &  0.15 \\
\ion{Sr}{II} &  2.92 & 1  &  1.55  &$-1.37$  &  $-$    \\
\ion{Y}{II}  &  2.21 & 2  &  1.08  &$-1.13$  &  0.34  \\
\ion{Zr}{II}  &  2.62 & 1  &  2.16  &$-0.46$ & $-$    \\
\ion{Ba}{II} &  2.17 & 1  &  1.14  &$-1.03$ & $-$   \\   
  \hline
\end{tabular}
\end{table}

\begin{table*}
\caption{Atmospheric abundances for the Star 1118\relax in Pal12 as obtained from our analysis of the HIRES  spectra analysed by \citet{C04}
compared with the original analysis of \citet{C04}.
}
\label{abu1118}
\centering
\begin{tabular}{l r r r r r r r r r r} 
\hline\hline                 
 & &\multicolumn{4}{c}{This paper}&\multicolumn{4}{c}{Cohen (2004)}\\
 & &\multicolumn{4}{c}{\teff = 4033 }&\multicolumn{4}{c}{\teff = 4000}\\
 & &\multicolumn{4}{c}{\logg = 0.74 }&\multicolumn{4}{c}{\logg = 0.84}\\
\hline\hline                 
Element & A(X)$_\odot$ &N(lines) & A(X) &  [X/H] & $\sigma$  &N(lines) & A(X) &  [X/H] & $\sigma$\\    
 & &\multicolumn{4}{c}{This paper}&\multicolumn{4}{c}{Cohen (2004)}\\
\hline
\ion{Na}{I}  &  6.30&  4 & 4.91& $-1.39$& 0.05 & 4 & 5.01 & --1.29 & 0.16\\
\ion{Mg}{I}  &  7.54&  1 & 6.89& $-0.65$&  $-$ & 2 & 6.85 & --0.69 & 0.11\\   
\ion{Si}{I}  &  7.52& 16 & 6.60& $-0.92$& 0.18 & 13& 6.78 & --0.74 & 0.16\\
\ion{Ca}{I}  &  6.33&  9 & 5.43& $-0.90$& 0.05 & 14& 5.31 & --1.02 & 0.22\\
\ion{Sc}{I}  &  3.10&  1 & 1.97& $-1.13$& $-$  & 6 & 2.25 & --0.85 & 0.16\\
\ion{Sc}{II} &  3.10&  6 & 2.11& $-0.99$& 0.18 &\\
\ion{Ti}{I}  &  4.90& 46 & 4.01& $-0.89$& 0.11 &  28& 4.05 & --0.85 & 0.16\\
\ion{Ti}{II} &  4.90& 15 & 3.91& $-0.99$& 0.29 &  5 & 4.19 & --0.71  & 0.13\\
\ion{V}{I}   &  4.00& 22 & 2.93& $-1.07$& 0.23 &  9 & 2.79 & --1.21  & 0.20\\
\ion{Cr}{I}  &  5.64& 19 & 4.63& $-1.01$& 0.13 &  6 & 4.89 & --0.75  & 0.23\\
\ion{Cr}{II} &  5.64&  3 & 4.64& $-1.00$& 0.19 & \\
\ion{Mn}{I}  &  5.37&  3 & 4.22& $-1.15$& 0.08 &  2 & 4.31 & --1.06  & 0.09\\
\ion{Fe}{I}  &  7.52& 171& 6.49& $-1.03$& 0.19 & 131 & 6.72 & --0.80 & 0.23 \\
\ion{Fe}{II} &  7.52&  7 & 6.62& $-0.90$& 0.15 & 9   & 6.81 & --0.71 & 0.19\\
\ion{Co}{I}  &  4.92& 17 & 3.72& $-1.20$& 0.21 & 4   & 3.80 & --1.12 & 0.28\\
\ion{Ni}{I}  &  6.23& 55 & 5.02& $-1.21$& 0.20 & 22  & 5.24 & --0.99 & 0.17\\
\ion{Cu}{I}  &  4.21&  2 & 2.46& $-1.75$& 0.13 & 2   & 3.55 & --0.66 & 0.16\\   
\ion{Zn}{I}  &  4.62&  2 & 3.01& $-1.61$& 0.08 & 2   & 3.16 & --1.46 & 0.16\\
\ion{Y}{II}  &  2.21&  9 & 0.92& $-1.29$& 0.25 & 2   & 0.84 & --1.37 & 0.52\\
\ion{Zr}{I}  &  2.62&  8 & 1.55& $-1.07$& 0.23 & 3   & 1.52 & -1.10  & 0.06\\   
\ion{Zr}{II} &  2.62&  1 & 2.40& $-0.22$&  $-$ \\   
\ion{Mo}{I}  &  1.92&  3 & 1.39& $-0.53$& 0.39 \\   
\ion{La}{II} &  1.14& 12 & 0.46& $-0.68$& 0.18 & 2 & 0.55 & --0.59 & 0.01\\
\ion{Ce}{II} &  1.61&  3 & 0.40& $-1.21$& 0.26 \\
\ion{Pr}{II} &  0.76&  1 &-0.13& $-0.89$&  $-$ \\   
\ion{Nd}{II} &  1.45& 31 & 0.75& $-0.70$& 0.35 &  2 & 1.02 & --0.43 & 0.26\\
\ion{Sm}{II} &  1.00&  7 & 0.22& $-0.78$& 0.34 \\
\ion{Eu}{II} &  0.52&  1 & 0.65& $+0.13$& $-$  & 1  & 0.30 &  -0.22 &    \\
  \hline
\end{tabular}
\end{table*}

  \begin{figure}
   \centering
   \resizebox{7.5cm}{!}{\includegraphics[clip=true]{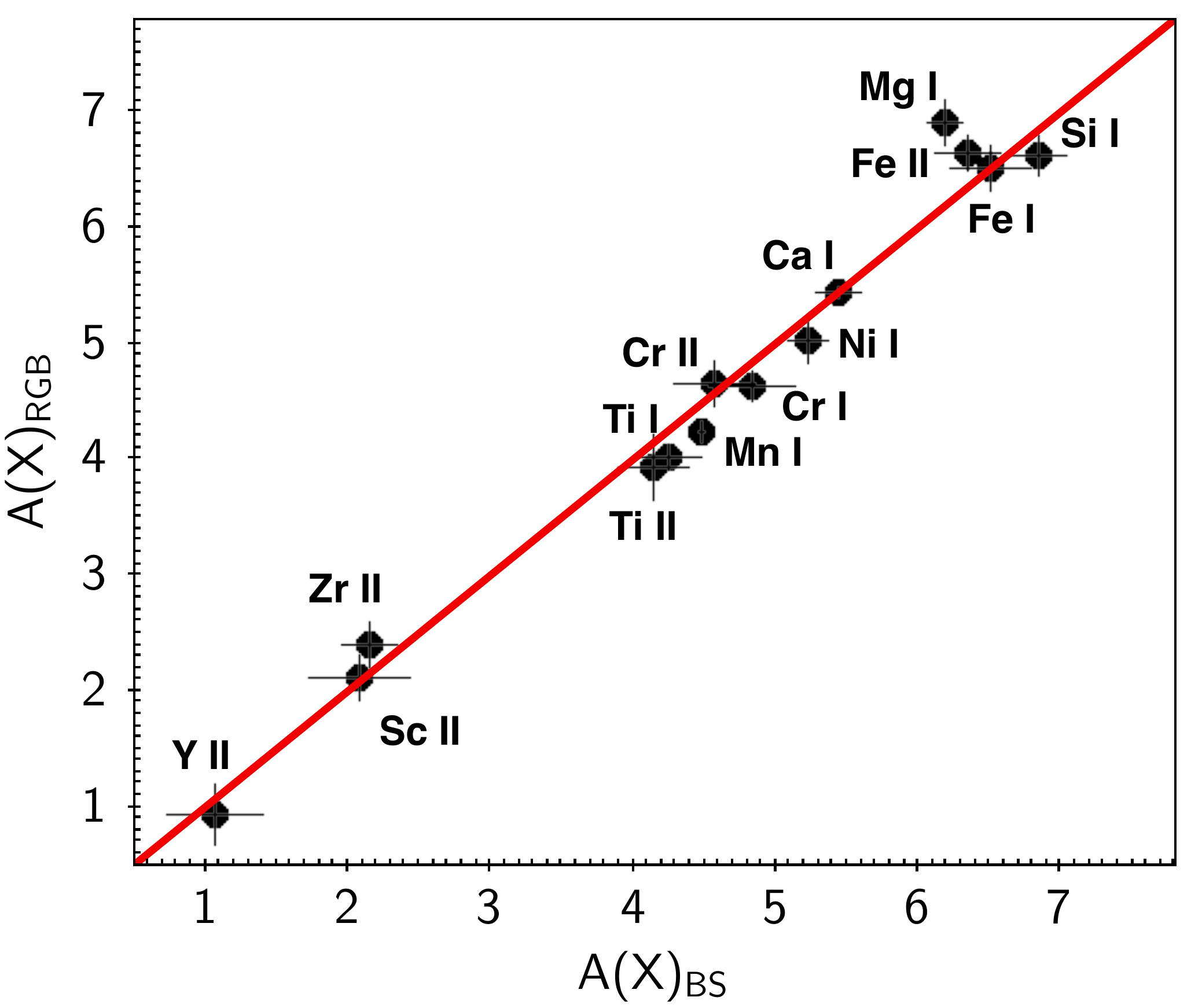}}
   \caption{Comparison of the abundances of the Blue Straggler 1336 with those
of the Red Giant Branch star 1118.}
   \label{BS_RGB}
   \end{figure}  

\section{Comparison with the abundances in a Red Giant Branch star}

In order to compare the chemical composition of the BS star 1336 with that of other stars in 
the cluster we retrieved from the Keck Archive the two spectra of the RGB star 1118  obtained by J. Cohen and whose analysis has been published in \citet{C04}.
The spectra were co-added and processed with MyGIsFOS \citep{SCBD14},using
a grid of synthetic spectra adequate for this giant,  based on ATLAS 12 model atmospheres and the same atomic and molecular data input as for the grid used to analyse the BS S1336.

We derived effective temperature and surface gravity from a Parsec isochrone \citep{parsec} and Gaia EDR3 photometry, as done for the BS star. The adopted parameters are \teff = 4033\,K and log g = 0.74, very close to the values derived by \citet{C04} for this star.
We also adopted the same microturbulent velocity of 1.8\,\kms as \citet{C04}.

The results of our  analysis are summarised in Table\,\ref{abu1118}.
Our abundances for this star are generally in good
agreement with  those of \citet{C04}. 

One strongly discrepant element is the Cu abundance (-1.09 dex difference between Cohen and our analysis), 
however this can be understood since in our analysis we have taken into account Hyperfine Splitting and Isotopic structure of the \ion{Cu}{I} lines. 
This causes the line to de-saturate, thus producing for the same equivalent width a 
much lower abundances. Our Cu abundance is based  on the 578.212\,nm line and
the 510.554\,nm line like \citet{C04}.  It is worth noting as our value is much closer to the average Cu in Pal12  of \citet{C04}, who indeed indicated 
Hyperfine Splitting as the likely cause of the large scatter of Cu abundance in her sample. 

Finally, Fe differs by 0.23 dex, an amount that 
could seem quite large,  however note that our measurement and that
of \citet{C04} are in agreement to better than 1\,$\sigma$, thus fully consistent.
Our analysis provides as well reasonably close values for \ion{Fe}{I} and \ion{Fe}{II}. The difference with \citet{C04} can be attributed to the choice of the Fe lines measured and to the adopted log gf values.

The analysis of the giant star 1118 is useful to set the comparison between the BS abundance and the cluster abundance on the same scale.
In Fig.\,\ref{BS_RGB}  the abundances measured in the
Blue Straggler (BS) with those of the Red Giant Branch (RGB) star are compared.
For all elements the agreement is good or very good, with the exception
of Mg. 
Mg in the BS is 0.5\,dex lower than in the RGB star.
It is true that the abundance in the RGB star is based on only one line, yet
our Mg abundance is in excellent agreement with that of \citet{C04} (that is based on two lines) for this star and also with the average value of Pal12. The three Mg lines used in our BS spectrum provide very consistent abundances, so we are  confident that also the BS  Mg determination is robust. 

It is unlikely that the difference is due to differential NLTE effects in the BS and in the giant, according to the NLTE
computations of \citep{mashonkina13} and \citep{alexeeva}  a star with \teff  of $\sim 7000$ K is barely affected by NLTE for the lines we used, and similarly, for the metal poor, cool giant  Arcturus the difference between LTE and NLTE abundance is less than 0.1 dex.  
Still, It would  be desirable
to perform dedicated NLTE computations for this case, to  completely rule out this possibility.

Another minor discrepancy is that of Mn (0.27\,dex), which, however, could be due
to differential 3D NLTE effects between the BS and the RGB. 
This is suggested by the 3D NLTE computations
of \citet[][see their figure 17]{bergemann19}. 
Again specific computations for these stellar models would be helpful to confirm it. 
BS and of the giant are comparable, in spite the BS being more than 4 magnitudes fainter. 

\section{Discussion \& summary}
\label{resume}

The study of stars close to the TO of globular clusters has been proven to be fundamental to determine the nature of these fascinating objects. Chemical abundances are close to pristine in these stars, while they can change in red giants during the course of the stellar evolution; in addition, our spectroscopic analysis may suffer of different biases and limitations for stars in different regimes of gravity and effective temperature. We can therefore learn a lot from  the determination of the chemical composition of stars close to the TO, and we have analysed several elements in a Blue Straggler belonging to the globular cluster Pal 12, associated to the Sagittarius Galaxy. The star is very faint (G=19.01) and to obtain this we have used ESPRESSO in the 4UT mode configuration, that provides a 16m equivalent telescope.  
We have analysed the spectrum a Pal 12 giant  previously observed with Keck-HIRES to compare the abundances of the BS  with those of the Pal 12 cluster, derived with the same method and lines GFs. The single abundances have comparable internal uncertainties and agree very well between the two stars, showing that the BS shares for most elements the Pal 12 abundance. 
Only Mg shows a strong discrepancy, with a lower abundance in the BS by 0.5 dex. Although NLTE effects 
cannot be completely excluded, they are very unlikely according to present calculations for similar stars and it is  
possible that the BS is indeed Mg-poor. Since there is no theoretical 
expectation for this to happen, we  expect that the BS atmosphere is dominated by material processed in the Mg-Al cycle. Unfortunately we have no suitable  Al or Na  lines to compute their abundance and   verify this hypothesis. Low Mg could be produced either by the merging of two 'second population' stars, or by mass transfer from a close companion during the AGB phase. 
Lithium abundance could help to discriminate between different formation mechanisms of BS, but the 3$\sigma$ abundance of Li$<$3.1 we obtain, is too high to derive a conclusion. 

\section*{Acknowledgements}

LPA acknowledges the scientific hospitality of the Arcetri Observatory. 
This research has made use of the Keck Observatory Archive (KOA), which is operated by the W. M. Keck 
Observatory and the NASA Exoplanet Science Institute (NExScI), under contract with the National Aeronautics and Space Administration. This work has made use of data from the European Space Agency (ESA) mission
{\it Gaia} (\url{https://www.cosmos.esa.int/gaia}), processed by the {\it Gaia}
Data Processing and Analysis Consortium (DPAC,
\url{https://www.cosmos.esa.int/web/gaia/dpac/consortium}). Funding for the DPAC
has been provided by national institutions, in particular the institutions
participating in the {\it Gaia} Multilateral Agreement.
\section*{Data Availability}
The ESPRESSO spectra are publicly available in the ESO archive, the Keck spectra are available from Keck Observatpry Archive



\bibliographystyle{mnras}
\bibliography{references} 



\bsp	
\label{lastpage}
\end{document}